\documentclass[12pt,reqno]{amsart}

\usepackage{amsmath,amsfonts,amssymb,amsthm,bm,fleqn}
\usepackage{enumerate}
\renewenvironment{proof}{\vskip .2em\indent{\bf Proof.}\ \ }{\hfill$\Box$}

\theoremstyle{theorem}
\newtheorem{theorem}{\indent\bf Theorem}

\theoremstyle{remark}

\newtheorem{result}{\indent\bf Result}

\renewcommand{\theequation}{\arabic{section}.\arabic{equation}}
\def\suml{\sum\limits}
\def\sumin{\sum\limits_{i=1}^n}

\def\sumtn{\sum\limits_{t=1}^n}
\def\sumjm{\sum\limits_{j=1}^m}
\def\ge{\geqslant}
\def\le{\leqslant}
\def\RR{\mathbb{R}}
\def\cc#1{\{#1\}}
\def\pn{(p_1,\ldots,p_n)}
\def\pk{(p_1,\ldots,p_k)}
\def\qm{(q_1,\ldots,q_m)}
\def\rn{(r_1,\ldots,r_n)}
\def\rns{(r_1^*,\ldots,r_n^*)}

\def\rnsG{\rns\in\Gamma_n}
\def\rnG{\rn\in\Gamma_n}
\def\pnG{\pn\in\Gamma_n}
\def\qmG{\qm\in\Gamma_m}
\def\pkG{\pk\in\Gamma_k}
\usepackage{titlesec,soul}
\newcommand\allcaps[1]{\MakeUppercase{\caps{#1}}}
\titleformat{\section}[block]{\mbox{}\indent\bfseries}{\thesection.}{1em}{}
\titlespacing*{\section}{0pt}{18pt}{10pt}

\def\pnn#1{(p_1,\ldots,p_n;#1)}
\def\rnns#1{(r_1^*,\ldots,r_n^*;#1)}
\def\rnn#1{(r_1,\ldots,r_n;#1)}

\title[A~\MakeLowercase{sum form functional equation .\,.\,.}]{}

\author[P. Nath and D.K. Singh]{}

\begin{document}

\begin{center}
\textbf{\large A sum form functional equation on a closed domain\\[.25em] and its role in information theory}

\bigskip
{P. Nath and D.K. Singh (India)}

\end{center}

\begin{abstract}\baselineskip 18pt
This paper is devoted to finding the general solutions of the functional
equation
\begin{eqnarray*}
\sumin \sumjm h(p_iq_j)=\sumin h(p_i)+\sumjm k_j(q_j)+\lambda\sumin h(p_i)\sumjm k_j(q_j)
\end{eqnarray*}
valid for all complete probability distributions
$(p_1,\ldots,p_n)$, $(q_1,\ldots,q_m)$,\break $0\le p_i\le 1$, $0\le q_j\le 1$,
$i=1,\ldots,n$; $j=1,\ldots,m$, $\sumin p_i=1$, $\sumjm q_j=1$; $n\ge 3$,  $m\ge 3$ fixed integers;
$\lambda\in\RR$, $\lambda\neq 0$ and the mappings $h:I\to\RR$, $k_j:I\to\RR$,
$j=1,\ldots,m$; $I=[0,1]$, $\RR$ denoting the set of all real numbers. A~special case of the above functional equation
was treated earlier by L.~Losonczi and Gy.~Maksa.
\end{abstract}



\maketitle
\thispagestyle{empty}

\vfill\eject
\baselineskip 23.75pt

\section{Introduction}

\mbox{}\indent Let $\Gamma_n=\cc{(p_1,\ldots,p_n):0\le p_i\le 1, i=1,\ldots,n;\sumin p_i=1}$, $n=2,3,\ldots$ denote the set of all
discrete $n$-component complete probability distributions with nonnegative elements. Let $\RR$ denote the set of all real
numbers and
\begin{eqnarray*}
\Delta&=&\cc{(x,y):0\le x\le 1, \ 0\le y\le 1, \ 0\le x+y\le 1}, \ \text{the unit triangle}\,;\\
I&=&\cc{x\in\RR:0\le x\le 1}=[0,1]~; ~~ I_0=\cc{x\in\RR:0<x<1}\,.
\end{eqnarray*}

A mapping $a:I\to\RR$ is said to be additive on $I$ if
\begin{eqnarray*}
a(x+y)=a(x)+a(y)
\end{eqnarray*}
holds for all $(x,y)\in\Delta$. A mapping $A:\RR\to\RR$ is said to be additive on $\RR$ if
\begin{eqnarray}
A(x+y)=A(x)+A(y)
\end{eqnarray}
holds for all $x\in\RR$, $y\in\RR$.

It is known \cite{2} that every mapping $a:I\to\RR$, additive on the unit triangle $\Delta$, has a unique additive extension $A:\RR\to\RR$ in the
sense that $A$ satisfies the equation (1.1) for all $x\in\RR$, $y\in\RR$.

A mapping $M:I\to\RR$ is said to be multiplicative on $I$ if
\begin{eqnarray}
M(0)=0\\
M(1)=1
\end{eqnarray}
and
\begin{eqnarray}
M(pq)=M(p)\,M(q)
\end{eqnarray}
holds for all $p\in I_0$, $q\in I_0$.

The functional equation (see \cite{1})
\begin{eqnarray}
\sumin \sumjm f(p_iq_j)=\sumin f(p_i)+\sumjm f(q_j)+\lambda \sumin f(p_i)\sumjm f(q_j)
\end{eqnarray}
where $f:I\to\RR$, $(p_1,\ldots,p_n)\in \Gamma_n$, $(q_1,\ldots,q_m)\in\Gamma_m$,
$\lambda=2^{1-\alpha}-1\neq 0$ is useful in characterizing the entropy of degree $\alpha$
(see \cite{3}) defined as
\begin{eqnarray}
H_n^\alpha (p_1,\ldots,p_n)=(1-2^{1-\alpha})^{-1}\left(1-\sumin p_i^\alpha\right),
\end{eqnarray}
where $H_n^\alpha:\Gamma_n\to\RR$, $n=2,3,\ldots$ and $0^\alpha:=0$, $\alpha\neq 1$, $\alpha\in\RR$.
For $\lambda\in\RR$, $\lambda\neq 0$, the general solutions of (1.5), for fixed integers
$n\ge 3$, $m\ge 3$
and all $\pn\in\Gamma_n$, $\qm\in\Gamma_m$ have been obtained in \cite{6}. A generalization of (1.5)
is the following functional equation (see \cite{5})
\begin{eqnarray}
\sumin \sumjm f_{ij}(p_iq_j)=\sumin h_i(p_i)+\sumjm k_j(q_j)+\lambda \sumin h_i(p_i)\sumjm k_j(q_j)
\end{eqnarray}
with $f_{ij}:I\to\RR$, $h_i:I\to\RR$, $k_j:I\to\RR$, $i=1,\ldots,n$; $j=1,\ldots,m$. For fixed integers $n\ge 3$,
$m\ge 3$ and all $\pn\in\Gamma_n$, $\qm\in\Gamma_m$, the measurable (in the sense of Lebesgue) solutions of (1.7)
have been obtained in (see \cite{5}, Theorem 6 on p-69) but it seems that the general solutions of (1.7), for fixed integers
$n\ge 3$, $m\ge 3$ and all $\pn\in\Gamma_n$, $\qm\in\Gamma_m$ are still not known. As mentioned in \cite{5}, equations like
(1.7) arise while characterizing measures of information concerned with two probability distributions.
In this paper, we study the equation
\begin{eqnarray}
\sumin \sumjm h(p_iq_j)=\sumin h(p_i)+\sumjm k_j(q_j)+\lambda\sumin h(p_i)\sumjm k_j(q_j)
\end{eqnarray}
where $h:I\to\RR$, $k_j:I\to\RR$, $j=1,\ldots,m$; $\lambda\in\RR$, $\lambda\neq 0$ and $n\ge 3$, $m\ge 3$ are fixed
integers. The functional equation (1.8) is a special case of (1.7).

If we define $f:I\to\RR$ and $g_j:I\to\RR$, $j=1,\ldots,m$ as (with $\lambda \neq0$)
\begin{eqnarray}
f(x)=x+\lambda\, h(x)\quad\text{and}\quad g_j(x)=x+\lambda\, k_j(x)
\end{eqnarray}
for all $x\in I$, then (1.8) reduces to the functional equation
\begin{eqnarray}
\sumin \sumjm f(p_iq_j)=\sumin f(p_i)\sumjm g_j(q_j)\,.
\end{eqnarray}
Also, (1.9) and $(1.10)$ yield (1.8). Thus, if the general solutions of (1.10),
for fixed integers $n\ge 3$, $m\ge 3$ and all $\pn\in\Gamma_n$, $\qm\in\Gamma_m$ are known; the~corresponding general
solutions of (1.8), for fixed integers $n\ge 3$, $m\ge 3$ and all $\pnG$, $\qmG$ can be determined with the aid of
(1.9).

We would like to mention that, on open domain, namely when $f:I_0\to\RR$,
$h:I_0\to\RR$,
$g_j:I_0\to\RR$,
$k_j:I_0\to\RR$,
$j=1,\ldots,m$, the general
solutions of (1.8) and (1.10) for fixed integers $n\ge 3$, $m\ge 3$ and all $\pnG$, $\qmG$ have been found in \cite{4}.
The object of this paper is to determine the general solutions of (1.8) and (1.10), on the closed domain, namely when
$f:I\to\RR$, $h:I\to\RR$, $g_j:I\to\RR$, $k_j:I\to\RR$, $j=1,\ldots,m$; for fixed integers $n\ge 3$, $m\ge 3$ and all
$\pnG$, $\qmG$. While investigating these solutions, the functional equation
\begin{eqnarray}
\sumin \sumjm \varphi(p_iq_j)=\sumin \varphi(p_i)\sumjm \varphi(q_j)+m(n-1)\,\varphi(0)\sumin \varphi(p_i)
\end{eqnarray}
arises with $\varphi:I\to\RR$, $n\ge 3$, $m\ge 3$ fixed integers and $\pnG$, $\qmG$.

To deal with equations (1.8), (1.10) and (1.11), we need the results and methods from \cite{5} and \cite{6}.

\section{Some preliminary results}
\setcounter{equation}{0}

\mbox{}\indent We require the following two results in sections 3 and 4.
\def\sumik{\suml_{i=1}^k}

\mbox{}
\vskip-4.5em\mbox{}
\begin{result} \cite{6}.
Let $k\ge 3$ be a fixed integer and $c$ be a given constant. Suppose that a mapping $\psi:I\to\RR$ satisfies the functional
equation
\begin{eqnarray}
\sumik \psi(p_i)=c
\end{eqnarray}
for all $\pkG$. Then there exists an additive mapping $B:\RR\to\RR$ such that
\begin{eqnarray}
\psi(p)=B(p)-\dfrac{1}{k}\,B(1)+\dfrac{c}{k}
\end{eqnarray}
for all $p\in I$.
\end{result}

\baselineskip 24pt
\begin{result} \cite{5}.
If the mappings $\psi_j:I\to\RR$, $j=1,\ldots,m$ satisfy the functional equation
\begin{eqnarray}
\sumjm \psi_j(q_j)=0
\end{eqnarray}
for an arbitrary but fixed integer $m\ge 3$ and all $\qmG$, then there exists an additive mapping
$A:\RR\to\RR$ and the constants $c_j$ $(j=1,\ldots,m)$ such that
\begin{eqnarray}
\psi_j(p)=A(p)+c_j
\end{eqnarray}
for all $p\in I$ and $j=1,\ldots,m$ with
\begin{eqnarray}
A(1)+\sumjm c_j=0\,.
\end{eqnarray}
\end{result}

\section{The functional equation (1.11)}
\setcounter{equation}{0}
\mbox{}\indent In this section, we prove:

\begin{theorem}
Let $n\ge 3$, $m\ge 3$ be fixed integers and $\varphi:I\to\RR$ be a mapping which satisfies the functional equation
\emph{(1.11)} for all $\pnG$ and $\qmG$. Then $\varphi$ is of the form
\begin{eqnarray}
\varphi(p)=a(p)+\varphi(0)
\end{eqnarray}
where $a:\RR\to\RR$ is an additive mapping with
\begin{eqnarray}\left.
\begin{array}{lll}
{\rm(i)}&a(1)=-\,nm\,\varphi(0)&\text{if } \ \varphi(1)+(n-1)\,\varphi(0)\neq 1\\[.5em]
&\qquad{\rm or}&\\[.5em]
{\rm(ii)}&a(1)=1-n\,\varphi(0)&\text{if } \ \varphi(1)+(n-1)\,\varphi(0)= 1
\end{array}\right\}
\end{eqnarray}
or
\begin{eqnarray}
\varphi(p)=M(p)-B(p)
\end{eqnarray}
where $B:\RR\to\RR$ is an additive mapping with $B(1)=0$ and $M:I\to\RR$ is multiplicative on $I$
in the sense that it satisfies \emph{(1.2)}, \emph{(1.3)} and \emph{(1.4)} for all $p\in I_0$, $q\in I_0$.
\end{theorem}

\proof
Let us put $p_1=1$, $p_2=\ldots=p_n=0$ in (1.11). We obtain
\begin{eqnarray}
[\varphi(1)+(n-1)\,\varphi(0)-1]\left[\sumjm \varphi(q_j)+m(n-1)\,\varphi(0)\right]=0
\end{eqnarray}
for all $\qmG$. We divide our discussion into two cases.

\vskip.5em
Case 1. $\varphi(1)+(n-1)\,\varphi(0)-1\neq 0$.

In this case, (3.4) reduces to
\begin{eqnarray}
\sumjm \varphi(q_j)=-\,m(n-1)\,\varphi(0)
\end{eqnarray}
for all $\qmG$. By Result 1, there exists an additive mapping $a:\RR\to\RR$ such that
\begin{eqnarray}
\varphi(p)=a(p)-\dfrac{1}{m}\,a(1)-(n-1)\,\varphi(0)
\end{eqnarray}
for all $p\in I$. The substitution $p=0$, in (3.6), gives
\begin{eqnarray}
a(1)=-\,nm\,\varphi(0)\,.
\end{eqnarray}
From (3.6) and (3.7), (3.1) follows. Thus, we have obtained the solution (3.1) satisfying (i) in(3.2).

\vskip.5em
Case 2. $\varphi(1)+(n-1)\,\varphi(0)-1=0$.

Let us write (1.11) in the form
\begin{eqnarray}
\sumjm \left\{\sumin \varphi(p_iq_j)-\varphi(q_j)\sumin \varphi(p_i)-m(n-1)\varphi(0)q_j\sumin \varphi(p_i)\right\}=0\,.
\end{eqnarray}
Choose $\pnG$ and fix it. Define $\psi:\Gamma_n\times I\to\RR$ as
\begin{eqnarray}
\psi(p_1,\ldots,p_n;q)\!=\!
\sumin \varphi(p_iq)\!-\!\varphi(q)\!\sumin\!\varphi(p_i)-m(n-1)\varphi(0)q\!\sumin \varphi(p_i)
\end{eqnarray}
for all $q\in I$. By Result 1, there exists a mapping $A_1:\Gamma_n\times \RR\to\RR$, additive in the
second variable, such that
\begin{eqnarray}
\lefteqn{\sumin \varphi(p_iq)-\varphi(q)\sumin \varphi(p_i)-m(n-1)\,\varphi(0)\,q\sumin \varphi(p_i)}\nonumber\\
&=&A_1(p_1,\ldots,p_n;q)-\dfrac{1}{m}\,A_1(p_1,\ldots,p_n;1)
\end{eqnarray}
The substitution $q=0$, in (3.10), gives
\begin{eqnarray}
A_1\pnn{1}=m\,\varphi (0)\left[\sumin \varphi(p_i)-n\right]
\end{eqnarray}
as $A_1\pnn{0}$. From (3.10) and (3.11), we obtain
\begin{eqnarray}
\lefteqn{\sumin \varphi(p_iq)-\varphi(q)\sumin \varphi(p_i)-m(n-1)\,\varphi(0)\,q\sumin \varphi(p_i)}\nonumber\\
&=&A_1\pnn{q}-\varphi(0)\sumin \varphi(p_i)+n\,\varphi(0)\,.
\end{eqnarray}
Since $\pnG$ was chosen arbitrarily and then fixed, equation (3.12), indeed, holds for all $\pnG$ and all $q\in I$.

Let $x\in I$ and $\rnG$. Putting $q=xr_t$, $t=1,\ldots,n$ in (3.12); adding the resulting $n$ equations and using the
additivity of $A_1$ in the second variable, it follows that
\begin{eqnarray}
\lefteqn{\sumin\sumtn \varphi(xp_ir_t)-\sumtn \varphi(xr_t)\sumin \varphi(p_i)-m(n-1)\,\varphi(0)\,x\sumin \varphi(p_i)}\nonumber\\
&=&A_1\pnn{x}-n\,\varphi(0)\sumin \varphi(p_i)+n^2\,\varphi(0)\,.
\end{eqnarray}
Also, if we put $q=x$ and $p_i=r_i$, $i=1,\ldots,n$ in (3.12), we obtain
\begin{eqnarray}
\sumtn \varphi(xr_t)&=&\varphi(x)\sumtn \varphi(r_t)+m(n-1)\,\varphi(0)\,x\sumtn \varphi(r_t)\nonumber\\
&&+\,A_1\rnn{x}-\varphi(0)\sumtn \varphi(r_t)+n\,\varphi(0)\,.
\end{eqnarray}
From (3.13) and (3.14), we can obtain the equation
\begin{eqnarray}
\lefteqn{\sumin \sumtn \varphi(xp_ir_t)-[\varphi(x)+m(n-1)\,\varphi(0)\,x-\varphi(0)]}\nonumber\\
\lefteqn{\qquad\times\sumin \varphi(p_i)\sumtn \varphi(r_t)-n^2\,\varphi(0)}\nonumber\\
&&\qquad=A_1\pnn{x}+m(n-1)\,\varphi(0)\,x\sumin \varphi(p_i)\nonumber\\
&&\qquad\quad+\,A_1\rnn{x}\sumin \varphi(p_i)\,.
\end{eqnarray}
The symmetry of the left hand side of (3.15), in $p_i$ and $r_t$, $i=1,\ldots,n$; $t=1,\ldots,n$ gives rise to the equation
\begin{eqnarray*}
\lefteqn{A_1\pnn{x}+m(n-1)\,\varphi(0)\,x\sumin \varphi(p_i)+A_1\rnn{x}\sumin\varphi(p_i)}\nonumber\\
&=&A_1\rnn{x}+m(n-1)\,\varphi(0)\,x\sumtn \varphi(r_t)\nonumber\\
&&+\,A_1\pnn{x}\sumtn \varphi(r_t)
\end{eqnarray*}
which can be written in the form
\begin{eqnarray}
\lefteqn{[A_1\pnn{x}+m(n-1)\,\varphi(0)\,x]\left[\sumtn \varphi(r_t)-1\right]}\nonumber\\
&=&[A_1\rnn{x}+m(n-1)\,\varphi(0)\,x]\left[\sumin \varphi(p_i)-1\right]\,.
\end{eqnarray}
Equation (3.16) holds for all $\rnG$, $\pnG$ and all $x\in I$.

\vskip .5em
Subcase 2.1. $\sumtn \varphi(r_t)-1$ vanishes identically on $\Gamma_n$.

In this case,
\begin{eqnarray}
\sumtn \varphi(r_t)=1
\end{eqnarray}
holds for all $\rnG$. By Result 1, there exists an additive map $a:\RR\to\RR$ such that
\begin{eqnarray}
\varphi(p)=a(p)-\dfrac{1}{n}\,a(1)+\dfrac{1}{n}
\end{eqnarray}
for all $p\in I$. The substitution $p=0$, in (3.18), yields
\begin{eqnarray}
a(1)=1-n\,\varphi(0)\,.
\end{eqnarray}
From (3.18) and (3.19), (3.1) follows again. Thus, we have obtained the solution (3.1) satisfying (ii) in (3.2).

\vskip .5em
Subcase 2.2. $\sumtn\varphi(r_t)-1$ does not vanish identically on $\Gamma_n$.

Then, there exists a probability distribution $\rnsG$ such that
\begin{eqnarray}
\sumtn \varphi(r_t^*)-1\neq 0\,.
\end{eqnarray}
Setting $r_1=r_1^*,\ldots,r_n=r_n^*$ in (3.16), we obtain
\begin{eqnarray*}
\lefteqn{[A_1(p_1,\ldots,p_n;x)+m(n-1)\,\varphi(0)\,x]\left[\sumtn \varphi(r_t^*)-1\right]}\nonumber\\
&=&[A_1\rnns{x}+m(n-1)\,\varphi(0)\,x]\left[\sumin \varphi(p_i)-1\right]
\end{eqnarray*}
which gives, for all $x\in I$,
\begin{eqnarray}
A_1\pnn{x}=A(x)\left[\sumin \varphi(p_i)-1\right]-m(n-1)\,\varphi(0)\,x
\end{eqnarray}
where $A:\RR\to\RR$ is defined as
\begin{eqnarray}
A(y)=\left[\sumtn \varphi(r_t^*)-1\right]^{-1}[A_1\rnns{y}+m(n-1)\,\varphi(0)\,y]
\end{eqnarray}
for all $y\in\RR$. From (3.22), it is easy to verify that $A:\RR\to\RR$ is additive.
Also, from (3.11) (with $p_i=r_i^*$, $i=1,\ldots,n$) and (3.22), it is easy to derive
\begin{eqnarray}
A(1)=m\,\varphi(0)\,.
\end{eqnarray}
From (3.12) and (3.21), it follows that
\begin{eqnarray*}
\lefteqn{\sumin \varphi(p_iq)-\varphi(q)\sumin \varphi(p_i)-m(n-1)\,\varphi(0)\,q\sumin \varphi(p_i)}\\
&=&A(q)\sumin \varphi(p_i)-A(q)-m(n-1)\,\varphi(0)\,q-\varphi(0)\sumin \varphi(p_i)+n\,\varphi(0)
\end{eqnarray*}
which, upon using (3.23), gives
\begin{eqnarray}
\lefteqn{\sumin [\varphi(p_iq)+A(p_iq)+m(n-1)\,\varphi(0)\,p_iq-\varphi(0)]}\nonumber\\
\lefteqn{\quad-\,[\varphi(q)+A(q)+m(n-1)\,\varphi(0)\,q-\varphi(0)]}\nonumber\\
\lefteqn{\quad\times\sumin [\varphi(p_i)+A(p_i)+m(n-1)\,\varphi(0)\,p_i-\varphi(0)]}\nonumber\\
\lefteqn{\quad+\,[\varphi(q)+A(q)+m(n-1)\,\varphi(0)\,q-\varphi(0)]\,n(m-1)\,\varphi(0)=0\,.}
\end{eqnarray}
Define a mapping $B:\RR\to\RR$ as
\begin{eqnarray}
B(x)=A(x)+m(n-1)\,\varphi(0)\,x
\end{eqnarray}
for all $x\in\RR$. Then, $B:\RR\to\RR$ is additive. Moreover, from (3.23) and (3.25), it follows that
\begin{eqnarray}
B(1)=mn\,\varphi(0)\,.
\end{eqnarray}
With the help of (3.25), equation (3.24) can be written in the form
\begin{eqnarray}
\lefteqn{\sumin [\varphi(p_iq)+B(p_iq)-\varphi(0)]-[\varphi(q)+B(q)-\varphi(0)]}\nonumber\\
&&\times\sumin [\varphi(p_i)+B(p_i)-\varphi(0)]+\,n(m-1)\,\varphi(0)\,[\varphi(q)+B(q)-\varphi(0)]\nonumber\\
&&\qquad=0\,.
\end{eqnarray}
Define a mapping $M:I\to\RR$ as
\begin{eqnarray}
M(x)=\varphi(x)+B(x)-\varphi(0)
\end{eqnarray}
for all $x\in I$. Notice that though $B:\RR\to\RR$ but, in (3.28), we are
restricting its use only for all $x\in I$.

From (3.28), it is easy to see that (1.2) follows as $B(0)=0$. Also, from (3.26), (3.28) and the fact that
$\varphi(1)+(n-1)\,\varphi(0)=1$, it follows that
\begin{eqnarray}
M(1)=1+n(m-1)\,\varphi(0)\,.
\end{eqnarray}
Moreover, from (3.27) and (3.28), we get (for all $q\in I$)
\begin{eqnarray}
\sumin M(p_iq)-M(q)\sumin M(p_i)+n(m-1)\,\varphi(0)\,M(q)=0
\end{eqnarray}
which can be written in the form
\begin{eqnarray}
\sumin [M(p_iq)-M(q)M(p_i)+n(m-1)\,\varphi(0)\,M(q)\,p_i]=0\,.
\end{eqnarray}
By Result 1, there exists a mapping $E:\RR\times I\to\RR$, additive in the first variable,
such that\renewcommand{\theequation}{3.31a}
\begin{eqnarray}
M(pq)-M(p)M(q)+n(m-1)\,\varphi(0)\,M(q)\,p=E(p,q)-\dfrac{1}{n}\,E(1,q)
\end{eqnarray}
for all $p\in I$, $q\in I$. The substitution $p=0$ in (3.31a) and the use of (1.2) gives $E(1,q)=0$ for all
$q\in I$. Consequently, (3.31a) reduces to the equation
\renewcommand{\theequation}{\arabic{section}.\arabic{equation}}\setcounter{equation}{31}
\begin{eqnarray}
M(pq)-M(p)M(q)+n(m-1)\,\varphi(0)\,M(q)\,p=E(p,q)
\end{eqnarray}
for all $p\in I$, $q\in I$.

Now we prove that $n(m-1)\,\varphi(0)\neq 0$ is not possible.

If possible, suppose $n(m-1)\,\varphi(0)\neq 0$. Then, (3.29) gives $M(1)\neq 1$. Putting $q=1$ in (3.30), using (3.29) and the
fact that $M(1)-1\neq 0$, we get
\begin{eqnarray*}
\sumin M(p_i)=M(1)
\end{eqnarray*}
for all $\pnG$. By Result 1, there exists an additive mapping $A_2:\RR\to\RR$ such that
\begin{eqnarray}
M(p)=A_2(p)-\dfrac{1}{n}\,A_2(1)+\dfrac{1}{n}\,M(1)
\end{eqnarray}
for all $p\in I$. The substitution $p=0$, in (3.33), gives $A_2(1)=M(1)$ as $A_2(0)=0$ and $M(0)=0$. Hence
\begin{eqnarray*}
M(p)=A_2(p)
\end{eqnarray*}
for all $p\in I$. Thus $M$ is additive on $I$. Now, from (3.20), (3.26), (3.28), (3.29) and
the additivity of $M$ on $I$, we have
\begin{eqnarray*}
1\neq \sumtn \varphi(r_t^*)&=&M(1)-B(1)+n\,\varphi(0)\\
&=&1+n(m-1)\,\varphi(0)-nm\,\varphi(0)+n\,\varphi(0)=1
\end{eqnarray*}
a contradiction.

So, the only possibility is that $n(m-1)\,\varphi(0)=0$. Since $n\ge 3$, $m\ge 3$ are fixed integers, it follows that
$\varphi(0)=0$ and hence $\varphi(1)=1$. From this and (3.29), (1.3) follows. Since $\varphi(0)=0$,
equation (3.32) reduces to the equation
\begin{eqnarray}
M(pq)-M(p)\,M(q)=E(p,q)
\end{eqnarray}
for all $p\in I$, $q\in I$. The left hand side of (3.34) is symmetric in $p$ and $q$.
Hence $E(p,q)=E(q,p)$ for all $p\in I$, $q\in I$. Consequently, $E$ is also additive on
$I$ in the second variable.
We may assume that $E(p,\cdot)$ has been extended additively to the whole of $\RR$.

Let $p\in I$, $q\in I$, $r\in I$. From (3.34), we have
\begin{eqnarray}
E(pq,r)+M(r)\,E(p,q)&=&M(pqr)-M(p)\,M(q)\,M(r)\nonumber\\
&=&E(qr,p)+M(p)\,E(q,r)\,.
\end{eqnarray}

Now we prove that $E(p,q)=0$ for all $p\in I$, $q\in I$. If possible, suppose there exists a $p^*\in I$ and
a $q^*\in I$ such that $E(p^*,q^*)\neq 0$. Then, from (3.35)
\begin{eqnarray*}
M(r)=[E(p^*,q^*)]^{-1}\{E(q^*r,p^*)+M(p^*)E(q^*,r)-E(p^*q^*,r)\}
\end{eqnarray*}
from which it follows that $M$ is additive on $I$. Now, making use of (3.20), (3.26), (3.28), (1.3), the
additivity of $M$ and the fact that $\varphi(0)=0$, we obtain
\begin{eqnarray*}
1\neq \sumtn \varphi(r_t^*)=M(1)-B(1)+n\,\varphi(0)=1-mn\,\varphi(0)+n\,\varphi(0)=1
\end{eqnarray*}
a contradiction. Hence $E(p,q)=0$ for all $p\in I$, $q\in I$. Now, (3.34) reduces to the equation
\begin{eqnarray}
M(pq)=M(p)\,M(q)
\end{eqnarray}
for all $p\in I$, $q\in I$. From (3.36), (1.4) follows immediately for all $p\in I_0$, $q\in I_0$. Also,
since $\varphi(0)=0$, (3.28) reduces to (3.3) and (3.26) gives $B(1)=0$. This completes the proof of Theorem 1.\endproof

\section{The functional equation (1.10)}
\setcounter{equation}{0}
\mbox{}\indent In this section, we prove:

\begin{theorem}
Let $n\ge 3$, $m\ge 3$ be fixed integers and $f:I\to\RR$, $g_j:I\to\RR$, $j=1,\ldots,m$
be mappings which satisfy the functional equation \emph{(1.10)} for all $\pnG$ and $\qmG$. Then, any general solution of
\emph{(1.10)} is of the form
\begin{eqnarray}
f(p)=b(p)\,,\quad \text{$g_j$ any arbitrary real-valued mapping}
\end{eqnarray}
where $b:\RR\to\RR$ is an additive mapping with $b(1)=0$ or\topsep10pt
\begin{eqnarray}\left.
\begin{array}{l}
f(p)=[f(1)+(n-1)\,f(0)]\,a(p)+f(0)\\[.75em]
g_j(p)=a(p)+A^*(p)+g_j(0)
\end{array}\right\}
\end{eqnarray}
for all $j=1,\ldots,m$; with $a:\RR\to\RR$, $A^*:\RR\to\RR$ being additive maps and\mathindent0pt
\begin{eqnarray}\left.
\begin{array}{ll}
a(1)=1-\dfrac{n\,f(0)}{f(1)+(n-1)\,f(0)}\,,&f(1)+(n-1)\,f(0)\neq 0\\[1.25em]
A^*(1)=-\sumjm g_j(0)+\dfrac{nm\,f(0)}{f(1)+(n-1)\,f(0)}\,,&f(1)+(n-1)\,f(0)\neq 0
\end{array}\right\}
\end{eqnarray}
or\mathindent3em
\begin{eqnarray}\left.
\begin{array}{ll}
f(p)=f(1)[M(p)-B(p)]\,,&f(1)\neq 0\\[.5em]
g_j(p)=M(p)-B(p)+A^*(p)+g_j(0)
\end{array}\right\}
\end{eqnarray}
for all $j=1,\ldots,m$; with $B:\RR\to\RR$, $A^*:\RR\to\RR$ being additive maps, $B(1)=0$, $A^*(1)=-\suml_{j=1}^m g_j(0)$ and
$M:I\to\RR$ a multiplicative function in the sense that it satisfies \emph{(1.2)}, \emph{(1.3)} and \emph{(1.4)} for all $p\in I_0$,
$q\in I_0$.
\end{theorem}\topsep8pt

\proof
Put $p_1=1$, $p_2=\ldots=p_n=0$ in (1.10). We obtain
\begin{eqnarray}
\sumjm [f(q_j)+(n-1)f(0)]=[f(1)+(n-1)\,f(0)]\sumjm g_j(q_j)
\end{eqnarray}
for all $\qmG$.

\vskip .5em
Case 1.
\begin{eqnarray}
f(1)+(n-1)\,f(0)=0\,.
\end{eqnarray}
Then, (4.5) reduces to the equation
\begin{eqnarray}
\sumjm f(q_j)=-\,m(n-1)\,f(0)
\end{eqnarray}
valid for all $\qmG$. By Result 1, there exists an additive mapping $b:\RR\to\RR$ such that
\begin{eqnarray}
f(p)=b(p)-\dfrac{1}{m}\,b(1)-(n-1)\,f(0)
\end{eqnarray}
for all $p\in I$. The substitution $p=0$, in (4.8), gives
\begin{eqnarray}
b(1)=-\,nm\,f(0)\,.
\end{eqnarray}
From (4.8) and (4.9), it follows that
\begin{eqnarray}
f(p)=b(p)+f(0)
\end{eqnarray}
for all $p\in I$. From (4.6), (4.9) and (4.10), using the fact that $n\ge 3$, $m\ge 3$ are fixed integers, it follows that
\begin{eqnarray}
f(0)=0\,.
\end{eqnarray}
From (4.9) and (4.11), it follows that
\begin{eqnarray}
b(1)=0\,.
\end{eqnarray}
Also, (4.10) and (4.11) give
\begin{eqnarray}
f(p)=b(p)
\end{eqnarray}
for all $p\in I$. Also, from (1.10), (4.12), (4.13) and the additivity of $b:\RR\to\RR$, it follows that
$g_j$ can be any arbitrary real-valued mapping. Thus, we have
obtained the solution (4.1) in which $b$ satisfies (4.12).

\vskip.5em
Case 2. $f(1)+(n-1)\,f(0)\neq 0$.

In this case, (4.5) gives
\begin{eqnarray}
\sumjm g_j(q_j)=[f(1)+(n-1)\,f(0)]^{-1}\sumjm [f(q_j)+(n-1)\,f(0)]
\end{eqnarray}
which can be written in the form
\begin{eqnarray}
\sumjm \big\{g_j(q_j)-[f(1)+(n-1)\,f(0)]^{-1}[f(q_j)+(n-1)\,f(0)]\big\}=0\,.
\end{eqnarray}
This holds for all $\qmG$. By Result 2, there exists an additive mapping $A^*:\RR\to\RR$ and constants $c_j$
$(j=1,\ldots,m)$ such that
\begin{eqnarray}
g_j(p)-[f(1)+(n-1)\,f(0)]^{-1}[f(p)+(n-1)\,f(0)]=A^*(p)+c_j
\end{eqnarray}
with
\begin{eqnarray}
A^*(1)+\sumjm c_j=0\,.
\end{eqnarray}
The substitution $p=0$, in (4.16), gives
\begin{eqnarray}
c_j=g_j(0)-[f(1)+(n-1)f(0)]^{-1}\,nf(0)
\end{eqnarray}
for $j=1,\ldots,m$. From (4.17) and (4.18), we get $A^*(1)$ as mentioned in (4.3).

Also, from (4.16) and (4.18),
\begin{eqnarray}
g_j(p)=[f(1)+(n-1)\,f(0)]^{-1}[f(p)-f(0)]+A^*(p)+g_j(0)
\end{eqnarray}
for $j=1,\ldots,m$. Equation (4.19) tells us that if $f$ is known, then the corresponding form of $g_j(p)$,
$j=1,\ldots,m$, can be determined. To determine $f$, we eliminate $\sumjm g_j(q_j)$ from equations (1.10) and
(4.14). We obtain the equation
\begin{eqnarray}
\sumin \sumjm f(p_iq_j)&=&[f(1)+(n-1)\,f(0)]^{-1}\sumin f(p_i)\sumjm f(q_j)\nonumber\\
&&+\,[f(1)+(n-1)\,f(0)]^{-1}m(n-1)\,f(0)\sumin f(p_i)
\end{eqnarray}
valid for all $\pnG$ and $\qmG$.

Define a mapping $\varphi:I\to\RR$ as
\begin{eqnarray}
\varphi(x)=[f(1)+(n-1)\,f(0)]^{-1}\,f(x)
\end{eqnarray}
for all $x\in I$. Then (4.20) reduces to the functional (1.11) which also holds for all
$\pnG$ and $\qmG$. Moreover, $\varphi$ satisfies the condition
\begin{eqnarray}
\varphi(1)+(n-1)\,\varphi(0)=1\,.
\end{eqnarray}
Also, from (4.21),
\begin{eqnarray}
f(p)=[f(1)+(n-1)\,f(0)]\,\varphi(p)
\end{eqnarray}
for all $p\in I$ with $f(1)+(n-1)\,f(0)\neq 0$ and
\begin{eqnarray}
\varphi(0)=\dfrac{f(0)}{f(1)+(n-1)\,f(0)}\,.
\end{eqnarray}
From, (4.19), (4.23), (4.24), (3.1) and (ii) in (3.2), the forms of $f(p)$, $g_j(p)$ and $a(1)$, as mentioned
in (4.2) and (4.3), follow. Thus, we have obtained the
solution (4.2), of (1.10), subject to $a(1)$ and $A^*(1)$ as mentioned in (4.3).

The form of $\varphi$, given by (3.3), with $B(1)=0$, is also acceptable as in this case, $\varphi(0)=0$,
$\varphi(1)=1$ and hence $\varphi(1)+(n-1)\,\varphi(0)=1$.
Now, from (4.24), $f(0)=0$.
The solution (4.4), of (1.10), follows from (4.23), (4.19), (3.3), (1.2), (1.3), (1.4) and the fact that $f(0)=0$,
$B(1)=0$, $A^*(1)=-\sumjm g_j(0)$.
This completes the proof of Theorem 2.\endproof

\section{The functional equation (1.8)}
\setcounter{equation}{0}
\mbox{}\indent In this section, we prove:

\begin{theorem}
Let $n\ge 3$, $m\ge 3$ be fixed integers and $h:I\to\RR$, $k_j:I\to\RR$, $j=1,\ldots,m$ be mappings which
satisfy the functional equation \emph{(1.8)} for all $\pnG$ and $\qmG$ and $\lambda\neq 0$.
Then, any general solution of \emph{(1.8)} is of the form
\begin{eqnarray}
h(p)=\dfrac{1}{\lambda}\,[b(p)-p], \ \text{$k_j$ any arbitrary real-valued mapping}
\end{eqnarray}
where $b:\RR\to\RR$ is an additive mapping with $b(1)=0$ or
\begin{eqnarray}\left.
\begin{array}{l}
h(p)=\dfrac{1}{\lambda}\big\{[\lambda(h(1)+(n-1)\,h(0))+1]\,a(p)+\lambda\, h(0)-p\big\}\\[.75em]
k_j(p)=\dfrac{1}{\lambda}\big\{a(p)+A^*(p)+\lambda\, k_j(0)-p\big\}
\end{array}\right\}
\end{eqnarray}
for all $j=1,\ldots,m$; with $a:\RR\to\RR$, $A^*:\RR\to\RR$ being additive maps and
\begin{eqnarray}\left.
\begin{array}{ll}
a(1)=1-\dfrac{n\lambda\, h(0)}{\lambda(h(1)+(n-1)\,h(0))+1}\,,\\[1.25em]
\mbox{}\qquad\qquad \qquad\qquad\qquad\qquad  \lambda(h(1)+(n-1)\,h(0))+1\neq 0\\[1.25em]
A^*(1)=-\lambda\sumjm k_j(0)+\dfrac{nm\lambda\,h(0)}{\lambda(h(1)+(n-1)\,h(0))+1}\,,\\[1.25em]
\mbox{}\qquad\qquad \qquad\qquad\qquad\qquad \lambda(h(1)+(n-1)\,h(0))+1\neq 0\,.
\end{array}\right\}
\end{eqnarray}
or
\begin{eqnarray}\left.
\begin{array}{ll}
h(p)=\dfrac{1}{\lambda}\,\big\{[\lambda\, h(1)+1][M(p)-B(p)]-p\big\}\,,\quad [\lambda\, h(1)+1]\neq 0\\[1.25em]
k_j(p)=\dfrac{1}{\lambda}\,\big\{M(p)-B(p)+A^*(p)+\lambda\, k_j(0)-p\big\}
\end{array}\right\}
\end{eqnarray}
with $B:\RR\to\RR$, $A^*:\RR\to\RR$ being additive maps such that
\begin{eqnarray}
B(1)=0\,, \quad A^*(1)=-\lambda \sumjm k_j(0)
\end{eqnarray}
and $M:I\to\RR$ a multiplicative
function in the sense that it satisfies \emph{(1.2)}, \emph{(1.3)} and \emph{(1.4)} for all $p\in I_0$, $q\in I_0$.
\end{theorem}

\proof
Let us write (1.8) in the form
\begin{eqnarray}
\sumin \sumjm [\lambda\, h(p_iq_j)+p_iq_j]=
\sumin [\lambda\, h(p_i)+p_i]\sumjm [\lambda \,k_j(q_j)+q_j]\,.
\end{eqnarray}

Define the mappings $f:I\to\RR$ and $g_j:I\to\RR$,
$j=1,\ldots,m$ (with $\lambda\neq 0$), as in (1.9), for all $x\in I$. Then,
(5.6) reduces to the functional equation (1.10) whose respective solutions are given by (4.1); (4.2)
subject to the condition
(4.3); and (4.4) subject to $B(1)=0$, $A^*(1)=-\sumjm g_j(0)$; in which $b:\RR\to\RR$, $a:\RR\to\RR$, $A^*:\RR\to\RR$,
$B:\RR\to\RR$ are all additive functions and $M:[0,1]\to\RR$ is a multiplicative function.
Now, making use of (1.9) along with (4.1); (4.2) subject to (4.3); and (4.4) subject to $B(1)=0$ and
$A^*(1)=-\sumjm g_j(0)$; the required solutions (5.1); (5.2) subject to (5.3); and (5.4) subject to (5.5);
follow respectively.~\endproof

\titleformat{\section}[block]{\centering\bfseries}{\thesection.}{1em}{\rm\allcaps}

\normalsize

\vskip 3em
\noindent\begin{tabular}{ p{3.2in}}
{\bf Prem Nath}          \\
Department of Mathematics   \\
University of Delhi         \\
Delhi 110 007       \\
India \\
\texttt{E-mail:\,pnathmaths@gmail.com}\\
  
\end{tabular}
\vskip 3em
\noindent\begin{tabular}{ p{3.2in}}
{\bf Dhiraj Kumar Singh}          \\
Department of Mathematics   \\
Zakir Husain Delhi College\\
(University of Delhi)         \\
Jawaharlal Nehru Marg \\
Delhi 110 002       \\
India \\
\texttt{E-mail:\,dhiraj426@rediffmail.com}\\
  \end{tabular}

\baselineskip 16pt
\begin{thebibliography}{99}\itemsep=6pt

\bibitem{1}
\textbf{M. Behara and P. Nath.}
Additive and non-additive entropies of finite measurable partitions,
\emph{Probability and Information Theory II}, Lecture Notes in Math., 296, Berlin-Heidelberg-New York, 1973, 102--138.

\bibitem{2}
\textbf{Z. Dar\'oczy and L. Losonczi.}
\"Uber die Erweiterung der auf einer Punktmenge additiven Funktionen,
\emph{Publ. Math.} (Debrecen), 14 (1967), 239--245.

\bibitem{3}
\textbf{J. Havrda and F. Charvat.}
Quantification method of classification process, concept of structural $\alpha$-entropy,
\emph{Kybernetika} (Prague), 3 (1967), 30--35.

\bibitem{4}
\textbf{PL. Kannappan and P.K. Sahoo.}
On the general solution of a functional equation connected to sum from information measures on open domain-VI,
\emph{Radovi Matematicki}, 8 (1992), 231--239.

\bibitem{5}
\textbf{L. Losonczi.}
Functional equations of sum form,
\emph{Publ. Math.} (Debrecen), 32 (1985), 57--71.

\bibitem{6}
\textbf{L. Losonczi and Gy. Maksa.}
On some functional equations of the information theory,
\emph{Acta Math. Acad. Sci. Hung.}, 39 (1982), 73--82.

\end{thebibliography}
\end{document}